\documentclass[aps,twocolumn,showpacs,byrevtex]{revtex4}
\usepackage{times}
\usepackage{amsmath}
\usepackage{graphicx}
\usepackage{color}
\newcommand{\y}{Y(4260)}
\newcommand{\yy}{Y(4200)}
\newcommand{\z}{Z_c(3900)}
\newcommand{\x}{X(3872)}
\newcommand{\xx}{X(3915)}
\newcommand{\xxx}{X(3960)}
\newcommand{\pp}{\pi^+\pi^-}

\newcommand{\LL}{\ell^+\ell^-}
\newcommand{\EE}{e^+e^-}

\newcommand{\GG}{\gamma\gamma}
\newcommand{\etap}{\eta^\prime}
\newcommand{\psip}{\psi(3686)}

\newcommand{\jpsi}{J/\psi}
\newcommand{\piz}{\pi^0}

\newcommand{\ppjpsi}{\pi^+\pi^-J/\psi}
\newcommand{\wjpsi}{\omega J/\psi}
\newcommand{\chicz}{\chi_{c0}}

\def\Journal#1#2#3#4{{#1} {\bf #2}, #3 (#4)}

\def\PLB{Phys. Lett. B}
\def\PRL{Phys. Rev. Lett.}
\def\PRD{Phys. Rev. D}

\def\EPJC{Eur. Phys. J. C}

\begin{document}

\title{\boldmath Study of $e^+e^- \to \gamma \omega J/\psi$ and Observation of $X(3872) \to \omega J/\psi$}
\author{
M.~Ablikim$^{1}$, M.~N.~Achasov$^{10,d}$, P.~Adlarson$^{59}$, S. ~Ahmed$^{15}$, M.~Albrecht$^{4}$, M.~Alekseev$^{58A,58C}$, A.~Amoroso$^{58A,58C}$, F.~F.~An$^{1}$, Q.~An$^{55,43}$, Y.~Bai$^{42}$
% ...
\\
\vspace{0.2cm}
(BESIII Collaboration)\\
\vspace{0.2cm} {\it
$^{1}$ Institute of High Energy Physics, Beijing 100049, People's Republic of China\\
% ...
$^{4}$ Bochum Ruhr-University, D-44780 Bochum, Germany\\
% ...
$^{10}$ G.I. Budker Institute of Nuclear Physics SB RAS (BINP), Novosibirsk 630090, Russia\\
% ...
$^{15}$ Helmholtz Institute Mainz, Johann-Joachim-Becher-Weg 45, D-55099 Mainz, Germany\\
% ...
$^{42}$ Southeast University, Nanjing 211100, People's Republic of China\\
$^{43}$ State Key Laboratory of Particle Detection and Electronics, Beijing 100049, Hefei 230026, People's Republic of China\\
% ...
$^{55}$ University of Science and Technology of China, Hefei 230026, People's Republic of China\\
% ...
$^{58}$ (A)University of Turin, I-10125, Turin, Italy; (B)University of Eastern Piedmont, I-15121, Alessandria, Italy; (C)INFN, I-10125, Turin, Italy\\
$^{59}$ Uppsala University, Box 516, SE-75120 Uppsala, Sweden\\
% ...
$^{d}$ Also at the Novosibirsk State University, Novosibirsk, 630090, Russia\\
% ...
}
}

\date{\today}

\begin{abstract}

We study the $\EE\to\gamma\wjpsi$ process using 11.6~fb$^{-1}$ $e^+ e^-$ 
annihilation data taken at center-of-mass energies from 
$\sqrt{s}=4.008$~GeV to 4.600~GeV with the BESIII detector 
at the BEPCII storage ring.
The $\x$ resonance is observed for the first time in the $\wjpsi$ system 
with a significance of more than $5\sigma$.
The relative decay ratio of $\x\to\wjpsi$ and $\ppjpsi$ is 
measured to be $\mathcal{R}=1.6^{+0.4}_{-0.3}\pm0.2$, where 
the first uncertainty is statistical and the second systematic (the same hereafter). 
The $\sqrt{s}$-dependent cross section of $\EE\to\gamma\x$ is
also measured and investigated, and it can be described by a single 
Breit-Wigner resonance, referred to as the $\yy$, 
with a mass of $4200.6^{+7.9}_{-13.3}\pm3.0~{\rm MeV}/c^2$
and a width of $115^{+38}_{-26}\pm12~{\rm MeV}$. 
In addition, to describe the $\wjpsi$ mass distribution above 3.9~GeV/$c^2$, we
need at least one additional Breit-Wigner resonance, labeled as $\xx$, in the fit.
The mass and width of the $\xx$ are determined.
The resonant parameters of the $\xx$ agree with those of the $Y(3940)$ in $B\to K\wjpsi$ and of
the $X(3915)$ in $\GG\to\wjpsi$ observed by the Belle and {\em BABAR} experiments within errors.

\end{abstract}


\maketitle

The $\x$ resonance was first observed by the Belle experiment~\cite{bellex},
and confirmed by the CDF~\cite{CDFx}, D0~\cite{D0x}, {\em BABAR}~\cite{babarx}, 
LHCb~\cite{LHCbx}, and BESIII Collaborations~\cite{bes3x}. Its unusual properties
do not accommodate with a charmonium state, and thus, the $\x$ resonance is
widely explained as an unconventional meson candidate~\cite{review}. Since the
$\x$ mass is near the $\bar{D^0}D^{*0}$  mass threshold, it is
often interpreted as a hadronic molecule by theoretical 
models~\cite{molecule}. The hadronic molecule model predicts that
the decay of $\x\to\wjpsi$ is sensitive to its internal structure, and 
a precise measurement of the decay rate would help to determine 
the ratio of various components that contribute to the $\x$ wave function.
While the decay $\x\to\ppjpsi$, where $\pp$ is found
to be dominated by a $\rho^0$~\cite{CDF-pp}, violates the isospin symmetry
in the strong interaction, the $\x\to\wjpsi$ decay process conserves isospin symmetry, and thus
such a decay provides an excellent metric for probing its isospin-violation effect. 
Previously, the Belle and {\em BABAR} Collaborations only reported less than $5\sigma$ evidences
for the $\x\to\wjpsi$ decay~\cite{wjpsi}. A solid observation
is still lacking and necessary for improved interpretation of this
first experimentally observed state potentially composed of four quarks.

The BESIII Collaboration recently reported evidence 
for the radiative transition $\y\to\gamma\x$ in $\x \to \pi^+\pi^- J/\psi$ mode~\cite{bes3x}.  A charged 
charmoniumlike state $\z$, which is a good candidate for a 
four-quark state~\cite{four-quark}, was observed near 
$\sqrt{s}=4.26$~GeV by BESIII~\cite{bes3-zc} and Belle~\cite{belle-zc},
and later confirmed with CLEO-c's data at $\sqrt{s}=4.17$~GeV~\cite{cleo-zc}.
All these observations show potential connections among the
 $\x$, $\y$ and $\z$ resonances, and strongly hint towards
a common underlying nature for them. At the moment,
more supportive experimental observation for the transition 
process $\y\to\gamma\x$ is needed to establish these
connections.

The $Y(3940)$ resonance was observed by the Belle Collaboration~\cite{belle-y3940}
and confirmed by the {\em BABAR} Collaboration~\cite{babar-y3940}
in $B\to K\wjpsi$. Later on, both Belle and
{\em BABAR} reported observations of the $\xx$ resonance in $\GG\to\wjpsi$
process~\cite{x3915}, and it was suggested to be the same resonance as the 
$Y(3940)$ by the Particle Data Group (PDG)~\cite{pdg}. 
The underlying nature of the $\xx$ is 
still unclear. It was once considered as a candidate for the $\chi_{c0}(2P)$ charmonium state.
However, such kind of assignment was challenged by a recent 
Belle observation~\cite{chic0p}. Other interpretations, such as a tetraquark~\cite{tetra-x3915}
or a hadronic molecule~\cite{molecule-x3915} are proposed for the $\xx$. Morever, a theoretical
calculation predicted a $1^{++}$ tetraquark with mass near 3.95~GeV/$c^2$~\cite{tetra-liuyr}.
To make the situation more clear, it is important to provide additional data on the $\xx$.

In this Letter, we report the study of the process $\EE \to \gamma\wjpsi$,
with $\jpsi\to\LL$ ($\ell=e,~\mu$) and $\omega\to\pp\piz(\piz\to\GG)$,
using data samples collected with the BESIII detector~\cite{Ablikim:2009aa}.
We search for the $\x$ and $\xx$ resonances in the $\wjpsi$ system and study
the $\sqrt{s}$-dependent production cross section, $\sigma[\EE\to\gamma\x]$. 
The $\EE$ center-of-mass (CM) energies of the data sets range from 
$\sqrt{s}=4.008$ to 4.600~GeV (c.f. Supplemental Material~\cite{xsec-CM}), with
a total integrated luminosity of about 11.6~fb$^{-1}$.

The BESIII detector is described in detail elsewhere~\cite{Ablikim:2009aa,etof}.
{\sc geant4}~\cite{geant} based Monte Carlo~(MC) simulation samples
are used to optimize the 
event selection criteria, determine the detection efficiency, and 
estimate backgrounds. For the signal process, we generate $\EE\to
\gamma\x/\xx \to \gamma\wjpsi$ MC events, with $\jpsi\to \LL$ ($\ell=e,~\mu$)
and $\omega\to\pp\piz(\piz\to\GG)$ at each CM energy corresponding to data. 
The $\x/\xx\to \wjpsi$ decay is described with the phase-space
model from {\sc evtgen}~\cite{evtgen}. 
Initial-state-radiation (ISR) is simulated with {\sc kkmc}~\cite{kkmc}.
The maximum ISR photon energy is set to correspond to the 
3.90~GeV/$c^2$ production threshold of the $\gamma\x$ system.
The final-state-radiation (FSR) from charged final-state particles
are handled with {\sc photos}~\cite{photos}.


Events with four charged tracks with net zero charge are
selected.  For each charged track, the polar angle in the
multilayer drift chamber 
must satisfy $|\cos\theta|<0.93$, and the
point of closest approach to the $\EE$ interaction point must be
within $\pm 10$~cm in the beam direction and within $1$~cm in the
plane perpendicular to the beam direction. Since the $\pi^\pm$ 
from $\omega$ decay and $\ell^\pm$ from $\jpsi$ decay
are kinematically well separated, charged tracks with momenta 
larger than 1.0~GeV/$c$ in the laboratory frame are assumed
to be $\ell^\pm$, and the ones with momenta less than 1.0~GeV/$c$ 
are assumed to be $\pi^\pm$. The energy deposition of charged 
tracks in the electromagnetic calorimeter (EMC) is used to
separate $e$ and $\mu$. For $\mu^\pm$ candidates, the deposited
energy in the EMC are required to be less than 0.35~GeV, while for
$e^\pm$, it is required to be larger than 1.1~GeV.

Showers identified as good photon candidates must satisfy fiducial and
shower-quality requirements. The minimum EMC energy is 25~MeV for
barrel showers ($|\cos\theta|<0.80$) and 50~MeV for end-cap
showers ($0.86<|\cos\theta|<0.92$). To eliminate showers produced
by charged particles, a photon must be separated by at least 20
degrees from any charged track in the EMC. The time information 
from the EMC is also used to suppress electronic noise and energy deposits
unrelated to the event.
At least three good photon candidates are required in each event.

To improve the momentum and energy resolutions and to
reduce backgrounds, a five-constraint (5C) kinematic fit is applied 
to an event with the hypothesis $\EE\to \gamma\pp \piz \LL$, 
which constrains the sum of four momentum of the final-state particles 
to the initial colliding beams, and the mass of two photon 
combinations to the $\piz$ world average mass~\cite{pdg}. 
The $\chi^2$ over number of degree of freedom (ndf) of the
kinematic fit is required to be less than $100/5$. When there are
ambiguities due to multi-combinations or multi-photon candidates in one event, 
we choose the combination with the smallest $\chi^2$.

Background events such as $\EE\to\pp\psip/\piz\piz\psip\to\pp\piz\piz\jpsi$
with one photon candidate missing would also pass the 
previously described event selection. To
remove these backgrounds, we require $|M^{\rm recoil}(\pp)-m[\psip]|>8$~MeV/$c^2$
and $|M(\ppjpsi)-m[\psip]|>7$~MeV/$c^2$, where 
$M^{\rm recoil}(\pp)=\sqrt{(P_{\EE}-P_{\pp})^2}$, and $m[\psip]$ is the
mass of the $\psip$ according to Ref.~\cite{pdg}. Other background events, such as 
$\EE\to\etap\jpsi\to\gamma\wjpsi$, have the same event topology 
as the signal. Their contribution can be effectively vetoed by rejecting events satisfying both 
$0.93<M(\gamma\omega)<0.97$~GeV/$c^2$ and $M(\omega\jpsi)>3.9$~GeV/$c^2$.

\begin{figure}
\begin{center}
\includegraphics[height=3cm]{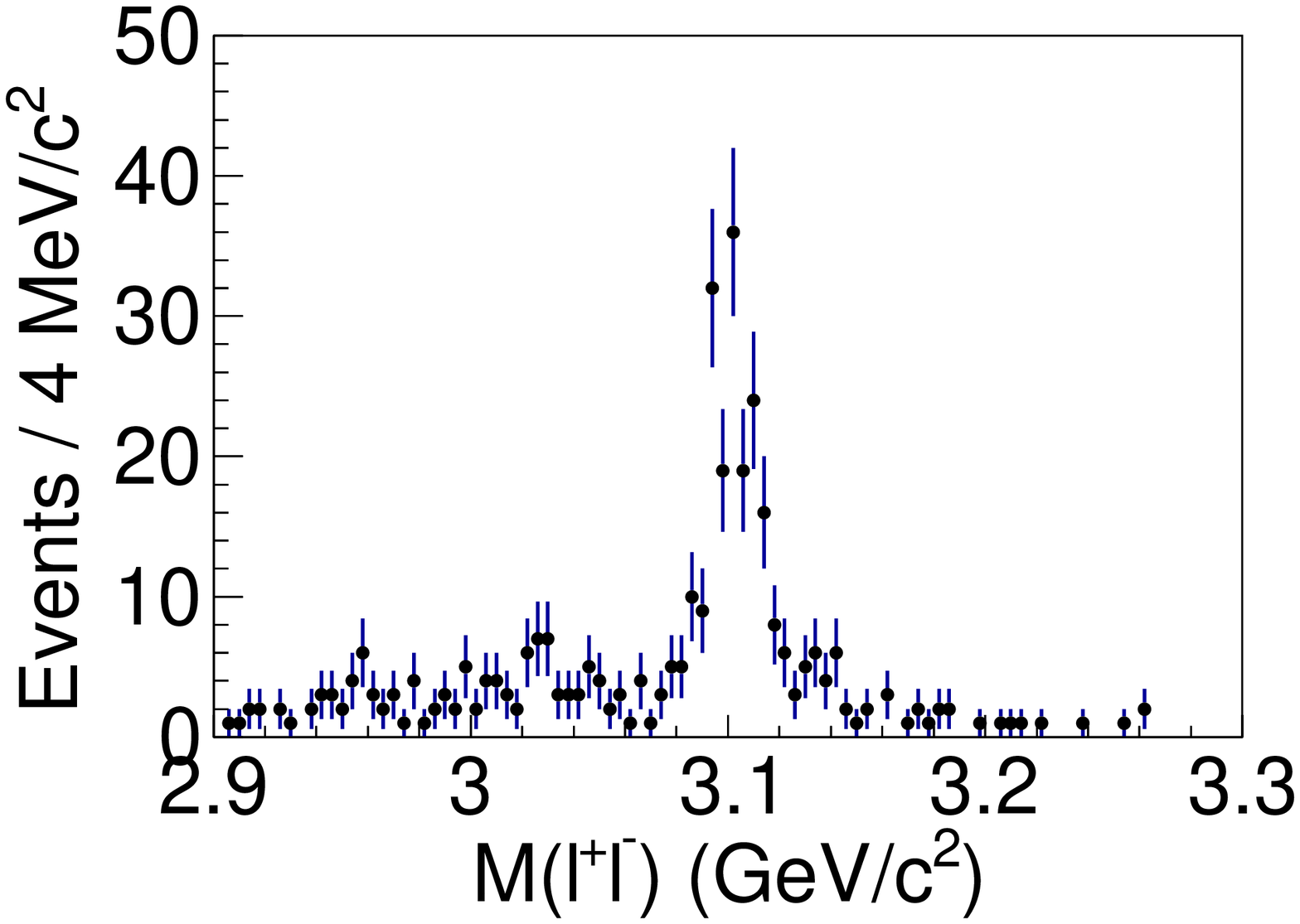}
\includegraphics[height=3cm]{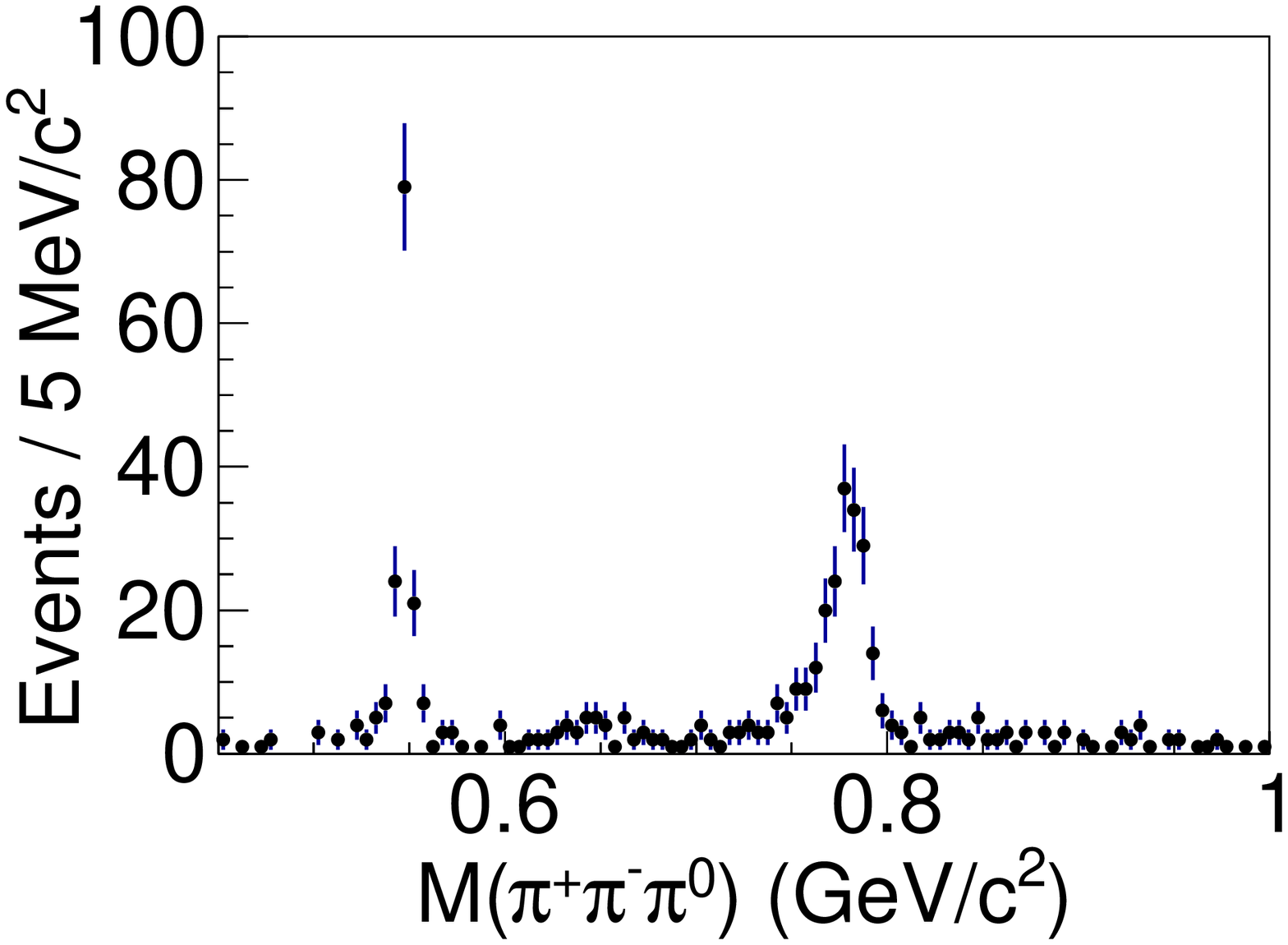}
\caption{The $M(\LL)$ and $M(\pp\piz)$ distributions from 
the full data sets.}
\label{fig-m2l-m3pi}
\end{center}
\end{figure}

After imposing the above requirements, clear peaks from $\jpsi$ and $\omega$
decays are seen in the $\LL$ and $\pp\piz$ invariant mass distributions, 
as shown in Fig.~\ref{fig-m2l-m3pi}. The $\eta$ peak in the
right panel of Fig.~\ref{fig-m2l-m3pi} comes from $\EE\to\eta\jpsi$ and 
$\gamma_{\rm ISR}\psip\to\gamma_{\rm ISR}\eta\jpsi$ processes.
To identify signal candidates that involve the $\jpsi$ resonances,
we select events within an invariant mass window of $3.07<M(\LL)<3.14$~GeV/$c^2$,
referred to as the $\jpsi$-mass window.
Non-$\jpsi$ background events are selected within the two sidebands
$2.97<M(\LL)<3.04$~GeV/$c^2$ or $3.17<M(\LL)<3.24$~GeV/$c^2$.

The difference between the mass of $\x$ and $\jpsi$~\cite{pdg} 
is about 775~MeV/$c^2$, which is slightly lower than the world average 
mass of the $\omega$. A consequence is an asymmetric $M(\pp\piz)$ 
distribution around the $\omega$ resonance, as can be seen 
in the right panel of Fig.~\ref{fig-m2l-m3pi}.
To accommodate for this effect, the $\omega$ mass window is defined as 
$0.72<M(\pp\piz)<0.81$~GeV/$c^2$, and its mass
sideband as $0.61<M(\pp\piz)<0.70$~GeV/$c^2$ or
$0.83<M(\pp\piz)<0.92$~GeV/$c^2$. We fitted both the 
$M(\LL)$ and $M(\pp\piz)$ distributions, and normalized the data of the sidebands
according to the fit results.

\begin{figure}
\begin{center}
\includegraphics[height=3cm]{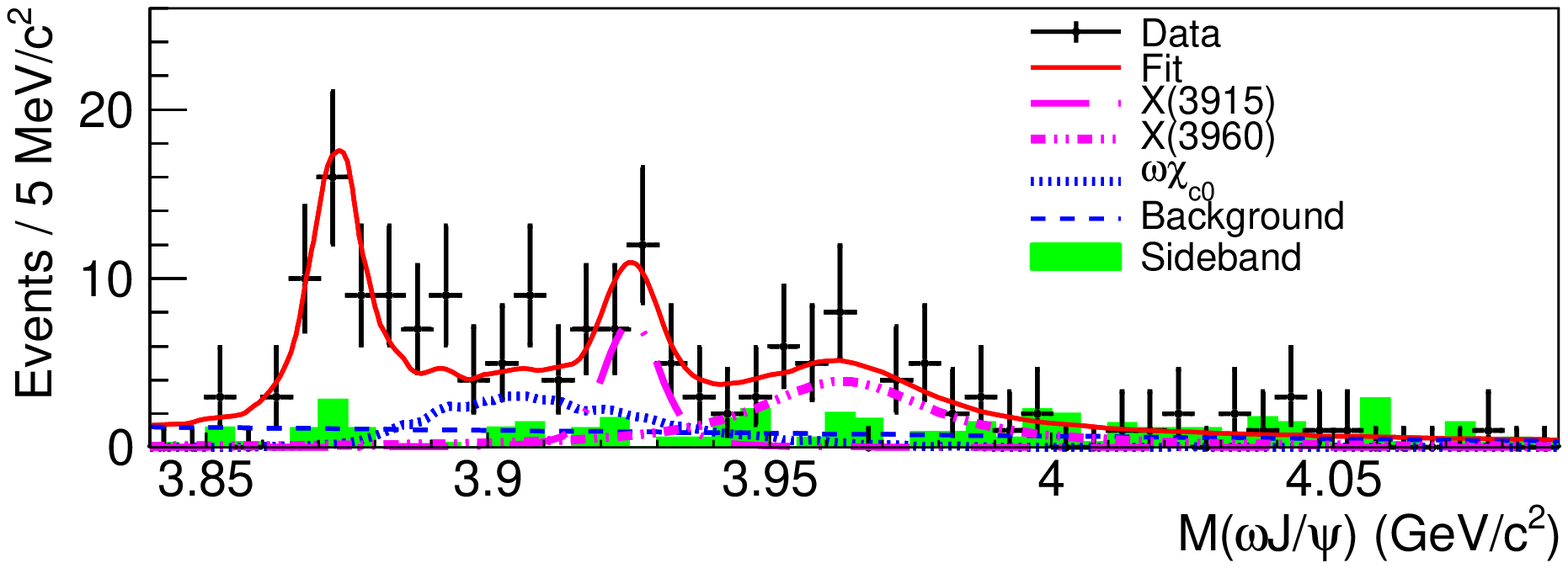}
\includegraphics[height=3cm]{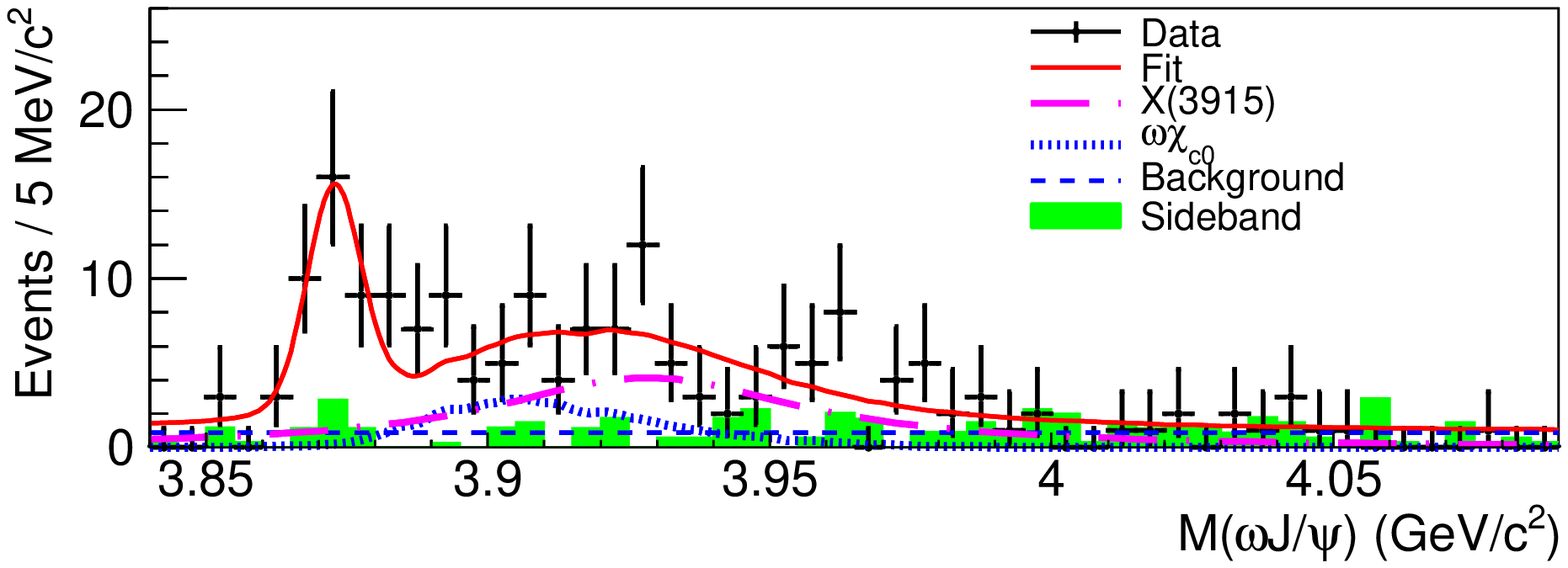}
\caption{The $M(\wjpsi)$ distribution with results of an unbinned 
maximum-likelihood fit to data including three BW resonances
(upper) and including two BW resonances (bottom) as signal.
Dots with error bars are data, the red solid curves show the total fit results,
the blue dotted curves are the MC simulated $\omega\chicz$ background component,
the blue dashed curves are the linear background component,
the pink dotted-dashed curves are the $\xx$ resonance, the pink double-dotted 
dashed curve is the $\xxx$ resonance, and the green shaded histograms
are the normalized contribution from the $\jpsi$- and $\omega$-mass sidebands.
}
\label{fig-fit-mx}
\end{center}
\end{figure}

Figure~\ref{fig-fit-mx} shows the $M(\wjpsi)$~\cite{x-mass} distribution 
from the full data set. 
A signal peak consistent with the $\x$ resonance is observed. 
In addition, there are evident structures above 3.9~GeV/$c^2$.
There are irreducible $\EE\to\omega\chicz$ background events
that produce a broad structure in the $M(\wjpsi)$ 
distribution. Such kind of background is 
well understood and can be reproduced by the MC simulation at BESIII~\cite{wcc0}.
Other possible backgrounds come from continuum events,
such as $\EE \to \gamma\omega\pp$, $\gamma\pp\piz\jpsi$, $\gamma\pp\piz\pp$ etc. 
They are estimated by analyzing the $\jpsi$ and $\omega$ mass sidebands data.

An unbinned maximum-likelihood 
fit is performed to the $M(\wjpsi)$ mass distribution. In the fit, we use
as the signal probability-density-function (PDF), the incoherent sum of three 
Breit-Wigner (BW) resonances (denoted as $\x$, $\xx$ and $\xxx$, respectively),
each convolved with a Gaussian resolution function. 
The $\x$ width is set to 1.2~MeV~\cite{pdg}.
The shape and yield of the $\EE\to\omega\chicz$ background component are fixed to the results
of the MC simulation. Contribution from other backgrounds is
parameterized as a linear shape. The upper panel of Fig.~\ref{fig-fit-mx} 
shows the fit results (numerical results are listed in Table~\ref{bw-para}), 
and the extracted $\x$ mass agrees with its world average value within errors.
The obtained $\x$ signal events yield is $N_{\rm sig}=45\pm9\pm3$.
The statistical significance of the $\x$
resonance is estimated to be $5.7\sigma$, by comparing the likelihood 
difference with or without the $\x$ in 
the fit, $\Delta(-2\ln\mathcal{L})=40.8$, and by taking the change of ndf 
($\Delta {\rm ndf}=3$) into account.
Possible systematic effects on the $\x$ signal significance, 
including background shape, $\omega\chicz$ background normalization,
$\x$ intrinsic width and mass resolution are investigated, and no
sign for a decreased $\x$ significance is observed.
The statistical significance of $\xx$ and 
$\xxx$ are estimated to be $3.1\sigma$ and $3.4\sigma$ only.

As an alternative choice, we fit the $M(\wjpsi)$ mass distribution
only with the $\x$ and $\xx$ resonances as signal PDF. The $\EE\to\omega\chicz$
background is handled in the same way as before. 
The contribution from other backgrounds is parameterized as 
a linear function and has been fixed to the result from fitting to
the data of the $\jpsi$- and $\omega$-mass sidebands.
The bottom panel of Fig.~\ref{fig-fit-mx} shows the fit results 
(c.f. Table~\ref{bw-para}), and the number of fitted $\x$
signal events is $N_{\rm sig}=40\pm8\pm3$.
The statistical significance of $\x$ is estimated to be $5.2\sigma$,
and found to be larger than $5.1\sigma$ after considering systematic
effects from $\omega\chicz$ and linear background normalization, $\x$ intrinsic width
and mass resolution.
The statistical significance of $\xx$ is estimated to be $6.9\sigma$.
We test the significance between these two fit scenarios, 
and find they only differ by $2.5\sigma$. 

\begin{table}
\begin{center}
\caption{The masses (in MeV/$c^2$) and widths (in MeV) of the $\x$,~$\xx$, and $\xxx$
resonances from the fit. The numbers in brackets represent the fit scenario without
the $\xxx$. The uncertainties are statistical only.} \label{bw-para}
\begin{tabular}{ccc}
  \hline\hline
   & Mass & Width \\
  \hline
  $\x$ & $3873.3 \pm 1.1~(3872.8\pm 1.2)$ & $1.2~(1.2)$ \\
  $\xx$ & $3926.4 \pm 2.2$~$(3932.6\pm8.7)$ & $3.8\pm 7.5$~$(59.7\pm15.5)$ \\
  $\xxx$ & $3963.7 \pm 5.5$ & $33.3\pm34.2$ \\
  \hline\hline
\end{tabular}
\end{center}
\end{table}

\begin{figure}
\begin{center}
\includegraphics[height=3.3cm]{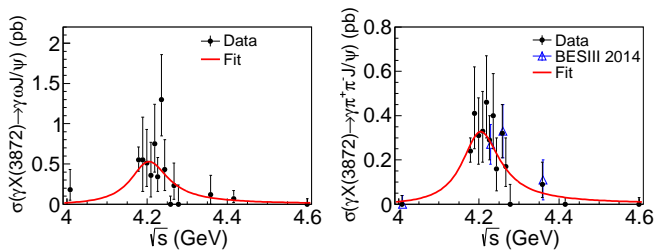}
\caption{The measured cross section of $\sigma[\EE\to\gamma\x]$ times the
branching fraction of $\x\to\wjpsi$ (left) and $\ppjpsi$ (right), and a simultaneous
fit to data with a single BW resonance.
Dots with error bars are data, the open triangles are an early measurement reported in
Ref.~\cite{bes3x}, and the red curves show the fit results.}
\label{fig-xsec}
\end{center}
\end{figure}

The production cross section of $\EE\to\gamma\x$ times the branching 
fraction $\mathcal{B}[\x \to \wjpsi]$ at each CM energy is calculated as
$\sigma\cdot\mathcal{B}[\x\to\wjpsi]=\frac{N^{\rm sig}}{\mathcal{L}\epsilon(1+\delta)\mathcal{B}}$,
where $N^{\rm sig}$ is the number of $\x$ signal events,
$\mathcal{L}$ is the integrated luminosity, $\epsilon$ is the detection
efficiency, $\mathcal{B}$ is the product of branching fractions for $\jpsi\to\LL$ 
and $\omega\to\pp\piz(\piz\to\GG)$, and $1+\delta$ is the ISR radiative
correction factor, which is calculated using the {\sc kkmc} program~\cite{kkmc}.
The ISR photon energy distribution is obtained by an iterative procedure using
the line shape $\sigma[\EE\to\gamma\x]$ measured in this study to replace the
default one of {\sc kkmc}.
The left panel of Fig.~\ref{fig-xsec} shows the measured 
$\sigma\cdot\mathcal{B}[\x\to\wjpsi]$. Using the same analysis method
as described in Ref.~\cite{bes3x} and the radiative correction factor in this study, 
$\sigma\cdot\mathcal{B}[\x\to\ppjpsi]$ 
is measured as well. Our result
agrees with and supersedes the earlier published BESIII measurement~\cite{bes3x},
as shown in the right panel of Fig.~\ref{fig-xsec}.
All the numerical results can be found in Supplemental Materials~\cite{xsec-CM}.

A simultaneous maximum-likelihood fit is performed to both the
$\sigma\cdot\mathcal{B}[\x\to\wjpsi]$ and the $\sigma\cdot\mathcal{B}[\x\to\ppjpsi]$ distributions.
We use a single BW resonance, denoted as $Y(4200)$, with
free mass and width as PDF. A free
parameter $\mathcal{R}=\frac{\mathcal{B}[\x\to\wjpsi]}{\mathcal{B}[\x\to\ppjpsi]}$ 
is used to describe the relative decay rate of $\x\to\wjpsi$ and $\ppjpsi$, 
which is common for every CM energy. The fit gives 
$M[\yy]=4200.6^{+7.9}_{-13.3}$~MeV/$c^2$,
$\Gamma[\yy]=115^{+38}_{-26}$~MeV, 
$\Gamma^{ee}\cdot\mathcal{B}[\yy\to\gamma\x]\cdot\mathcal{B}[\x\to\ppjpsi]=(4.5^{+1.1}_{-0.8})\times 10^{-2}$~eV 
and $\mathcal{R}=1.6^{+0.4}_{-0.3}$, where $\Gamma^{ee}$ 
is the electronic partial width of the $\yy$. Here, all the uncertainties are
statistical only.

The systematic uncertainty for $\x$, $\xx$, and $\xxx$ mass and width measurements
come from the uncertainties in the absolute mass scale, background and resolution effects.
The $\EE\to\gamma_{\rm ISR}\psip\to\gamma_{\rm ISR}\eta\jpsi$ 
events with the same event selection (except the $\omega$ mass window
is replaced by the $\eta$ mass window) are used as a control sample to
calibrate the mass scale. The measured $\psip$ mass is
$3685.4\pm0.4$~MeV/$c^2$, and the difference to the $\psip$ world
average mass is 0.8~MeV/$c^2$. Backgrounds are
varied from a linear shape to a second-order polynomial or by
$\pm 1\sigma$ for the linear component, and varied 
by $\pm 1\sigma$ for the $\omega\chicz$ component in the fit.
The differences in the mass and width measurements with
respect to the nominal results are taken as a systematic uncertainty. 
The systematic uncertainty of resolution is estimated
by varying the Gaussian parameters of the resolution response function 
by $\pm 1\sigma$ in the signal PDF.
In both fit scenarios (with and without the $\xxx$), the $\x$ mass 
difference 0.5~MeV/$c^2$ is taken as a systematic uncertainty 
due to the fit model.
All these contributions are summarized
in Table~\ref{sys-I}, and the total uncertainty is calculated
by adding the independent contributions in quadrature.

\begin{table}
\begin{center}
\caption{Summary of the systematic uncertainties for $\x$/$\xx$/$\xxx$ mass
and width measurements. The numbers in the brackets correspond to
the fit scenario with only the $\x$ and $\xx$ as signal PDF.} \label{sys-I}
\begin{tabular}{ccc}
  \hline\hline
  Source & Mass (MeV/$c^2$) & Width (MeV) \\
  \hline
  Absolute mass scale & 0.8/0.8~(0.8)/0.8 & -/-/- \\
  Background shape & 0.3/0.4~(4.5)/0.5 & -/2.5~(3.6)/8.3 \\
  Resolution & 0.0/0.8~(0.7)/0.8 & -/0.7~(0.3)/0.1 \\
  Fit model & 0.5/-/- & -/-/- \\
  \hline
  Total & 1.0/1.2~(4.7)/1.3 & -/2.6~(3.7)/8.3 \\
  \hline\hline
\end{tabular}
\end{center}
\end{table}


The systematic uncertainty for the $\EE\to\gamma\x$ cross section measurement mainly
comes from uncertainties in the luminosity measurements, detection efficiency, 
signal extraction, radiative correction and branching fractions. The integrated luminosities of 
each data set are measured with large-angle Bhabha scattering events, with an uncertainty 
of 1.0\%~\cite{lum}. The tracking efficiency is estimated to be 1\% per track
from a study of the control sample $\jpsi\to p\bar{p}\pp$. 
The uncertainty due to the photon reconstruction is studied using the $\jpsi\to\pp\piz$ events, 
and is found to be 1\% for the radiative photon~\cite{gamma-error}.
An additional systematic uncertainty of 1\% is assigned to the efficiency of $\pi^0$ reconstruction
by studying $\psip\to\piz\piz\jpsi$ and $\EE\to\omega\piz$ events. 
In our event selection, a 5C kinematic fit is used, and the systematic uncertainty 
related to the kinematic fit is estimated to be 0.8\% by using a helix correction 
method as discussed in Ref.~\cite{KF}. 

The number of $\x$ signal events is extracted by fitting the
$M(\wjpsi)$ distribution, and the difference between the two fit scenarios
is 9.5\%. 
The $\x$ intrinsic width is fixed to 1.2~MeV in the signal PDF. Varying the width
from 50~keV to 1.2~MeV results in a 5\% difference for the $\x$ signal yield. 
The systematic uncertainty of the $\omega\chicz$ background is estimated 
by varying the normalization by $\pm 1\sigma$, which will cause
a difference of 0.9\% in the $\x$ signal yield.
The remaining background is parameterized as a linear function. 
Varying the background shape from
linear to a second-order polynomial or the normalization by $\pm 1\sigma$ 
will cause a 3.1\% difference for the $\x$ signal yield. 

We iterate the cross section measurement until the value
of $(1+\delta)\epsilon$ changes by at most 1\% from the previous iteration,
and 1\% is taken as a systematic uncertainty due to ISR radiative correction.
The systematic uncertainty related to the $\jpsi$-mass window cut is 
1.6\%~\cite{bes3x}. The branching fraction uncertainties of $\jpsi\to\LL$,
$\omega\to\pp\piz$ and $\piz\to\GG$ are 0.6\%, 0.8\% and 0.04\%~\cite{pdg}, 
respectively.

The total systematic uncertainty is calculated to be 12.3\% by adding all
contributions in quadrature.


The systematic uncertainty for the $\yy$ parameters mainly comes from
the uncertainties related to the
$\EE$ CM energy measurement, the parameterization of the fit model,
and the cross section measurement. The CM energy of each data
set is measured with dimuon events, with $\pm 0.8$~MeV
uncertainty~\cite{ecm}. Such kind of common uncertainty will 
shift the $\yy$ line shape globally, and thus, propagate to the $\yy$ 
mass linearly. In the fit to cross section, the $\yy$ resonance is 
parameterized as a BW with a constant full width. We
also use a BW with a phase-space dependent full
width, $\Gamma\frac{\Phi(\sqrt{s})}{\Phi(M)}$, and the
difference is 2.8~MeV/$c^2$ for the mass, 12~MeV for the width,
and 6.5\% for $\Gamma^{ee}$. 
The cross section data measured in $\x\to\wjpsi$
and $\ppjpsi$ channels are fitted simultaneously. The common 
uncertainties of cross section measurements in both channels, 
including luminosity, tracking, photon detection, radiative correction,
kinematic fit, $\x$ intrinsic width, $\jpsi$ mass window, 
and $\jpsi\to\LL$ branching fraction, will propagate to 
$\Gamma^{ee}$ linearly, {\it i.e.} 6.9\%. 
The uncommon ones, including $\piz$, background,
fit model and $\omega\to\pp\piz(\piz\to\GG)$ branching fraction,
will affect the $\mathcal{R}$ measurement, and the total 
contribution is 10.9\%, by adding them in quadrature.


In summary, we have studied the $\EE\to\gamma\wjpsi$ process 
with 11.6~fb$^{-1}$ data at the BESIII experiment. For the first time, 
the $\x\to\wjpsi$ decay was firmly observed with more than 
$5\sigma$ significance, and the $\x$ mass was measured to
be $3873.3\pm1.1\pm1.0$~MeV/$c^2$. 
The relative decay ratio for $\x\to\wjpsi$ and $\ppjpsi$ is 
measured to be $\mathcal{R}=1.6^{+0.4}_{-0.3}\pm0.2$, which
agrees well with previous measurements within errors~\cite{wjpsi}.
These measurements provide important input for the hadronic
molecule interpretation for the $\x$ resonance~\cite{molecule}.

To describe the $M(\wjpsi)$ distribution above 3.9~GeV/$c^2$,
we need at least one additional BW resonance $\xx$. Its mass
and width are measured to be $3926.4\pm2.2\pm1.2$~MeV/$c^2$
and $3.8\pm7.5\pm2.6$~MeV; or $3932.6\pm8.7\pm4.7$~MeV/$c^2$
and $59.7\pm15.5\pm3.7$~MeV, depending on the fit models.

The $\EE\to\gamma\x$ production cross section is 
measured at the CM energies between 4.008 and 4.600~GeV~\cite{xsec-CM}.
We studied the $\sqrt{s}$-dependent cross section line shape of
$\EE\to\gamma\x$, and find it
can be described by a single BW resonance $\yy$. A simultaneous fit to the 
$\x\to\wjpsi$ and $\ppjpsi$ cross section data gives its mass
$M[\yy]=4200.6^{+7.9}_{-13.3} \pm 3.0$~MeV/$c^2$, and 
width $\Gamma[\yy]=115^{+38}_{-26} \pm 12$~MeV, 
which agree with the $\psi(4160)$~\cite{pdg} or the $Y(4220)$
observed by BESIII in $\ppjpsi$~\cite{bes3-ppjpsi} and 
$\pp h_c$~\cite{bes3-pphc} within errors. 
The measured $\EE\to\gamma\x$ cross section provides useful
information for the $D\bar{D^*}$ hadronic molecule calculation
as described in Ref.~\cite{zhaoq-model}.

The BESIII collaboration thanks the staff of BEPCII and the IHEP computing center for their strong support. This work is supported in part by National Key Basic Research Program of China under Contract No. 2015CB856700; National Natural Science Foundation of China (NSFC) under Contracts Nos. 11335008, 11425524, 11625523, 11635010, 11735014; the Chinese Academy of Sciences (CAS) Large-Scale Scientific Facility Program; the CAS Center for Excellence in Particle Physics (CCEPP); Joint Large-Scale Scientific Facility Funds of the NSFC and CAS under Contracts Nos. U1532257, U1532258, U1732263; CAS Key Research Program of Frontier Sciences under Contracts Nos. QYZDJ-SSW-SLH003, QYZDJ-SSW-SLH040; 100 Talents Program of CAS; INPAC and Shanghai Key Laboratory for Particle Physics and Cosmology; German Research Foundation DFG under Contract No. Collaborative Research Center CRC 1044; Istituto Nazionale di Fisica Nucleare, Italy; Koninklijke Nederlandse Akademie van Wetenschappen (KNAW) under Contract No. 530-4CDP03; Ministry of Development of Turkey under Contract No. DPT2006K-120470; National Science and Technology fund; The Knut and Alice Wallenberg Foundation (Sweden) under Contract No. 2016.0157; The Swedish Research Council; U. S. Department of Energy under Contracts Nos. DE-FG02-05ER41374, DE-SC-0010118, DE-SC-0012069; University of Groningen (RuG) and the Helmholtzzentrum fuer Schwerionenforschung GmbH (GSI), Darmstadt.


\end{document}